\documentclass[prd,twocolumn,eqsecnum,amsfonts,amssymb]{revtex4}
\pdfoutput=1
\usepackage{url}
\usepackage{hyperref} % see the file hyperref.cfg for options
\hypersetup{
	colorlinks=true,
	citecolor=,
	linkcolor=,
	filecolor=,      
	urlcolor=blue,
}
\usepackage{graphicx}
\usepackage{caption}
\usepackage{subcaption}
\usepackage{bm}

\newcommand{\beq}{\begin{equation}}
\newcommand{\eeq}{\end{equation}}
\newcommand{\beqs}{\begin{eqnarray}}
\newcommand{\eeqs}{\end{eqnarray}}

\newcommand{\gsim}{\mathrel{\raisebox{-
.6ex}{$\stackrel{\textstyle>}{\sim}$}}}

\begin{document}

\title{Fitting a Self-Interacting Dark Matter Model to Data Ranging 
From Satellite Galaxies to Galaxy Clusters}

\author{Sudhakantha Girmohanta$^{1,2}$ and Robert Shrock$^1$}

\affiliation{1 \ C. N. Yang Institute for Theoretical Physics and
Department of Physics and Astronomy, \\
Stony Brook University, Stony Brook, New York 11794, USA }

\affiliation{2\ Tsung-Dao Lee Institute and School of Physics and Astronomy,\\
Shanghai Jiao Tong University, 800 Dongchuan Road, Shanghai 200240, China}

\begin{abstract}

  We present a fit to observational data in an asymmetric self-interacting dark
  matter model using our recently calculated cross sections that incorporate
  both $t$-channel and $u$-channel exchanges in the scattering of identical
  particles.  We find good fits to the data ranging from dwarf galaxies to
  galaxy clusters, and equivalent relative velocities from $\sim 20$ km/sec to
  $\gsim 10^3$ km/s. We compare our results with previous fits that used only
  $t$-channel exchange contributions to the scattering. 

\end{abstract}

\maketitle

% section 1
\section{Introduction}
\label{intro_section}

There is strong evidence for dark matter (DM), comprising about 85 \% of the
matter in the universe.  Cold dark matter (CDM) can account for structures on
length scales larger than $\sim 10$
Mpc~\cite{primack,nfw,nfw2,kravtsov_primack,moore,fsw2020} (reviews include
\cite{jkg,binney_tremaine, bhs, strigari,
  lisanti_review,dm_history,buckley_peter_review}.)  However, problems have
been noted with fits to observational data on shorter length scales of $\sim
1-100$ kpc using early CDM simulations without baryon feedback
\cite{spergel_steinhardt,boylan_kolchin,boylan_kolchin2}.  These problems
included the prediction of greater density in the central region of galaxies
than was observed (the core-cusp problem), a greater number of dwarf satellite
galaxies than were seen (the missing satellite problem), and the so-called
``too big to fail'' problem pertaining to star formation in dwarf satellite
galaxies.  Models with self-interacting dark matter (SIDM) have been shown to
avoid these problems (some reviews include
\cite{zurek_adm_review,tulin_yu_review,adhikari_rmp}).  The extension of cold
dark matter $N$-body simulations to include baryon feedback can ameliorate
these problems with pure CDM simulations~\cite{springel, springel2,
  scannapieco, governato2012,zolotov2012,fire2015, fire2016, nfw_apostle, bullock_boylan_kolchin, peter,
  fbbb, chua_vogelsberger_hernquist,springel2019, fire2}. Nevertheless,
cosmological models with self-interacting dark matter (SIDM) are of
considerable interest in their own right and have been the subject of intensive
study \cite{zurek_adm_review,tulin_yu_review,adhikari_rmp,
  dssw,kusenko_steinhardt, mirror1, bullet_cluster_constraints, afsw,
  feng_kaplinghat_yu, buckley_fox, koda, vogelsberger_zavala_loeb,
  kouvaris2012, lin_yu_zurek, tulin_yu_zurek2013prl, tulin_yu_zurek2013, wise,
  cline,
  kusenko, frandsen_sarkar, elbert_bullock_peter, kaplinghat_tulin_yu_prl2016,
  slatyer_sommerfeld, valli_reassessment,ethos, kkpy,battaglieri, ksw, mirror2,
  robertson2017, robertson2018, valli_yu, lesgourgues_brinckmann_sidm,
  brinckmann_vogelsberger, vogelsberger_zavala_slatyer, robertson2019, rkky,
  essig_sidm, salucci, wechsler, apr, hayashi, kaplinghat2020, ccm, bondarenko,
  essig_loverde, tulin_velocity, slone_evolution,
  tulin_etal2021, ebisu, fischer2022a,ray,
  lisanti2022, bullock_kaplinghat_valli2022, eckert2022, fischer2022,
  nadav2022, yangyu2022,sidm}.

In the framework of a particle theory of dark matter, the rate of DM-DM
scatterings is given by $\Gamma = (\sigma/m_{\rm DM})v_{\rm rel}\rho_{_{\rm
    DM}}$, where $\sigma$, $m_{\rm DM}$, $v_{\rm rel}$, and $\rho_{_{\rm DM}}$
are the DM-DM scattering cross section, DM particle mass, relative velocity of
two colliding DM particles, and DM mass density, respectively. Fits to
observational data on the scale of $\sim 1-10$ kpc, with velocities $v_{\rm
  rel} \sim 20-200$ km/s, yield values $\sigma/m_{\rm DM} \sim 1$ cm$^2$/g,
while fits to observations of galaxy clusters on distance scales of several Mpc
and $v_{\rm rel} \sim O(10^3)$ km/s yield smaller values of $\sigma/m_{\rm DM}
\sim 0.1$ cm$^2$/g. This implies that viable SIDM models should have cross
sections that decrease as a function of $v_{\rm rel}$. This property can be
achieved in models in which DM particles, denoted $\chi$ here, interact via
exchange of a light (Lorentz scalar or vector) mediator field, generically 
denoted $\xi$. 

In models with asymmetric dark matter (ADM), after the number asymmetry is
established in the early universe, the DM self-interaction occurs via the
reaction
\beq
\chi + \chi \to \chi + \chi \ . 
\label{chichi_reaction}
\eeq
Because of the identical particles in the final state, a proper treatment
necessarily includes both the $t$-channel and the $u$-channel contributions to
the scattering amplitude.  In \cite{sidm}, we presented differential and
integrated cross sections for the reaction (\ref{chichi_reaction}) with both
the $t$-channel and $u$-channel terms included and discussed the differences
with respect to previous calculations that included only the $t$-channel
term. Identical-particle effects have also been noted in
\cite{ksw,yangyu2022} in a field-theoretic context and in 
\cite{cline,tulin_etal2021}
in the context of solutions of the Schr\"odinger
equation for potential scattering. An interesting
question raised by our work in \cite{sidm} is the following: how do the fits to
observational data change when one uses the cross section with both $t$-channel
and $u$-channel contributions to the scattering, as contrasted with 
previous fits that used only the $t$-channel contributions?  In the present
paper we address this question using the same observational data set that was
analyzed in \cite{kaplinghat_tulin_yu_prl2016}. 

% =================================================================

% section 2
\section{cross sections}
\label{sigma_section}

First, we review the basic properties of the SIDM model with asymmetric dark
matter that we used in \cite{sidm}. In this model, the dark matter particle
$\chi$ is a spin-1/2 Dirac fermion, and the mediator, $\xi$, is a real scalar,
$\xi=\phi$, or a vector, $\xi=V$. Both $\chi$ and $\xi$ are singlets under the
Standard Model (SM). For the version of the model with a real scalar mediator,
we take the $\chi$-$\phi$ interaction to be of Yukawa form, as described by the
interaction Lagrangian ${\cal L}_{\rm Yuk} = y_\chi [\bar\chi \chi]\phi$.  In
the version with a vector mediator, the DM fermion $\chi$ is assumed to be
charged under a U(1)$_V$ gauge symmetry with gauge field $V$ and gauge coupling
$g$.  Since only the product of the U(1)$_V$ charge of $\chi$ times $g$ occurs
in the covariant derivative in this theory, we may, without loss of generality,
take this charge to be unity and denote the product as $g_\chi$.  The
corresponding interaction Lagrangian is ${\cal L}_{\bar\chi\chi V} = g_\chi
[\bar\chi \gamma_\mu \chi]V^\mu$.  A Higgs-type mechanism is assumed to break
the U(1)$_V$ symmetry and give a mass $m_V$ to the gauge field $V$.  For
compact notation, we use the same symbol, $\alpha_\chi$, to denote
$y_\chi^2/(4\pi)$ for the case of a scalar mediator and $g_\chi^2/(4\pi)$ for
the case of a vector mediator.  We assume that the kinetic mixing of $V$ with
the SM hypercharge gauge boson is negligibly small. (For an example of how this
mixing can be suppressed in a DM model with specified ultraviolet physics, see,
e.g., \cite{dmled}.) In \cite{sidm} the illustrative set of values $m_\chi=5$
GeV, $m_\xi=5$ MeV, and $\alpha_\chi = 3 \times 10^{-4}$ was used.  Below we
will show that this choice is consistent with the fit to astronomical data that
we perform here.

We restrict to the case where $\alpha_\chi$ is small enough so that
lowest-order perturbation theory provides a reliable description of the
physics. As was shown in \cite{sidm}, the parameter choice used there satisfies
this restriction while simultaneously yielding sufficient depletion of the
$\bar\chi$ number density in the early universe to produce the assumed number
asymmetry in our ADM model. For further details on our model, we refer the
reader to \cite{sidm}.

The amplitude for the reaction (\ref{chichi_reaction}) 
is ${\cal M} = {\cal M}^{(t)} - 
{\cal M}^{(u)}$, where ${\cal M}^{(t)}$ and ${\cal M}^{(u)}$ are the 
$t$-channel and $u$-channel contributions and the minus sign embodies the 
effect of interchange of identical fermions in the final state. 
Let us define a prefactor $\sigma_0$ and dimensionless ratio 
$r$ as 
\beq
\sigma_0 = \frac{\alpha_\chi^2 m_\chi^2}{m_\xi^4} \ , \quad 
r = \bigg ( \frac{\beta_{\rm rel} m_\chi}{m_\xi} \bigg )^2 \ .
\label{sigr}
\eeq
where $\beta_{\rm rel} = v_{\rm rel}/c$.  For all the relevant data, the values
of $v_{\rm rel}$ are nonrelativistic (NR). In \cite{sidm} we calculated the
differential cross section in the center-of-mass (CM), 
$d\sigma_{\rm CM}/d\Omega$, for the reaction
(\ref{chichi_reaction}) with both scalar and vector mediators in the regime
where the Born approximation is valid. In the NR limit relevant to fitting
data, the results for the scalar and vector mediators are equal and are
\cite{sidm}
\begin{widetext}
\beqs
&& \bigg ( \frac{d\sigma}{d\Omega} \bigg )_{\rm CM,NR} = {\sigma_0} \bigg [
\frac{1}{(1+r\sin^2(\theta/2))^2} +
\frac{1}{(1+r\cos^2(\theta/2))^2} -
\frac{1}{(1+r\sin^2(\theta/2))(1+r\cos^2(\theta/2))} \bigg ] \ .  \cr\cr
&& 
\label{dsigma}
\eeqs
\end{widetext}
The terms on the right-hand side of Eq. (\ref{dsigma}) are from $|{\cal
  M}^{(t)}|^2$, $|{\cal M}^{(u)}|^2$, and $[{\cal M}^{(t) *} {\cal M}^{(u)}
+{\cal M}^{(u) *} {\cal M}^{(t)}]$, respectively. The angular integrals of
these terms are correspondingly denoted as $\sigma^{(t)}$, $\sigma^{(u)}$, and
$\sigma^{(tu)}$. Because of the identical particles in the final state, a
scattering event in which a scattered $\chi$ particle emerges at angle $\theta$
is indistingishable from one in which a scattered $\chi$ emerges at angle
$\pi-\theta$. The total cross section for the reaction (\ref{chichi_reaction})
thus involves a symmetry factor of $1/2$ to compensate for the double-counting
involved in the integration over the range $\theta \in [0,\pi]$:
\beq
\sigma = \frac{1}{2} \int d\Omega \,
\bigg ( \frac{d\sigma}{d\Omega} \bigg )_{\rm CM} \ .
\label{sigma_eq}
\eeq
Owing to the symmetry $\left( \frac{d\sigma}{d\Omega} \right)_{\rm CM}(\theta) =
\left( \frac{d\sigma}{d\Omega} \right)_{\rm CM}(\pi-\theta)$, this is
equivalent to a polar angle integration from 0 to $\pi/2$. 

To describe the thermalization effects of DM-DM scattering, cross
sections that give greater weight to large-angle scattering have also
been used in fits to data. These include the transfer ($T$) cross
section $d\sigma_T/d\Omega = (1-\cos \theta)(d\sigma/d\Omega)_{\rm
  CM}$ and the viscosity ($V$) cross section, $d\sigma_V/d\Omega =
(1-\cos^2 \theta)(d\sigma/d\Omega)_{\rm CM}$.  These have the
respective weighting factors $w_T(\theta)=1-\cos\theta$ and
$w_V(\theta) = 1-\cos^2 \theta$, as indicated.
Ref. \cite{tulin_yu_zurek2013} suggested the use of the viscosity
cross section $\sigma_V$ for studies of SIDM thermalization effects,
and recently, Ref. \cite{yangyu2022} finds that $\sigma_V$ provides a
very good description of thermalization effect of SIDM scattering.
However, since $\sigma_T$ has been used in a number of past fits to
observational data, we include results for it here for completeness.
We obtained the integrated cross sections (given as Eqs. (4.30) and
(4.38) in \cite{sidm})
\beq
\sigma = \sigma_T = 
4\pi \sigma_0 \bigg [ \frac{1}{1+r} - \frac{\ln(1+r)}{r(2+r)}
  \bigg ] 
\label{sigma}
\eeq
and
\beqs
\sigma_V &=& \frac{8\pi\sigma_0}{r^2}\bigg [
  -5 + \frac{2(5+5r+r^2)\ln(1+r)}{(2+r)r} \bigg ] \ . \cr\cr 
&& 
\label{sigma_v}
\eeqs
For a given weighting factor, the integrals of the terms
$(d\sigma/d\Omega)_{\rm CM}$ were denoted $\sigma^{(t)}$,
$\sigma^{(u)}$, and $\sigma^{(tu)}$ and analytic expressions for these
were given for $\sigma = \sigma_T$ and $\sigma_V$ in our previous work
\cite{sidm}.  We note that (when one includes both $t$-channel and
$u$-channel contributions) since the weighting factor $w_T(\theta) =
(1-\cos\theta)$ has no net effect in suppressing contributions from
scattering events that do not produce thermalization, it follows that
$\sigma_T$ could overestimate the thermalization effect from SIDM
self-scattering.

In the literature, in the same NR Born regime for the reaction
(\ref{chichi_reaction}) a formula was used for the differential cross section
of reaction (\ref{chichi_reaction}) that implicitly assumed that the colliding
particles were distinguishable (e.g., Eq. (5) in \cite{tulin_yu_review}),
namely (in our notation)
\beq
\bigg ( \frac{d\sigma}{d\Omega} \bigg )_{\rm CM,t} = \frac{\sigma_0}
{(1+r\sin^2(\theta/2))^2} \ . 
\label{dsig_lit}
\eeq
The integrals $\sigma = \int d\Omega \, w(\theta)(d\sigma/d\Omega)_{\rm CM}$ 
with the respective weighting factors (again assuming distinguishable 
particles in reaction (\ref{chichi_reaction}) ) yielded
results for $\sigma$, $\sigma_T$, and $\sigma_V$, in particular, the following
result for $\sigma_T$ (Eq. (5) in \cite{feng_kaplinghat_yu}, denoted FKY, 
which is the same as Eq. (A1) in \cite{tulin_yu_zurek2013prl}, denoted TYZ and 
Eq. (6) in \cite{tulin_yu_zurek2013})
\beq
\sigma_{T,FKY,TYZ} = \frac{8 \pi \sigma_0}{r}
\bigg[- \frac{1}{1+r} + \frac{\ln(1+r)}{r} \bigg] \ ,
\label{sigma_t_lit}
\eeq
In \cite{sidm} we showed that 
\beq
\sigma_{T,FKY,TYZ} = 2\sigma^{(t)} \ , 
\label{sigt_tyz_versus_our_sigt}
\eeq
where $\sigma^{(t)}$ was given as Eq. (4.27) in \cite{sidm}), and the factor of
2 difference is due to the fact that the correct calculation divides by $1/2$
to take account of the identical particles in the final state.  For brevity, we
denote $\sigma_{T,FKY,TYZ} \equiv \sigma_{T,{\rm lit.}}$ (lit. = cited
literature).

Following the same procedure as for $\sigma_{T,lit.}$, 
from Eq. (\ref{dsig_lit}), one would get
\beqs 
\sigma_{V,{\rm lit}.} = \frac{16\pi\sigma_0}{r^2}\bigg [ -2 +
(2+r)\frac{\ln(1+r)}{r} \bigg ] \ .
\label{sigma_v_lit}
\eeqs
This is twice as large as the term $\sigma_V^{(t)}$ that entered into the full
viscosity cross section $\sigma_V =
\sigma_V^{(t)}+\sigma_V^{(u)}+\sigma_V^{(tu)}$ that we calculated in
\cite{sidm} (see Eqs. (4.36) and (4.38) in \cite{sidm}).  The origin of the
factor 2 difference is again that when one takes account of the identical
particles in the final state, the angular integral (\ref{sigma_eq}) involves a
factor of $1/2$. In addition to \cite{sidm}, more recent studies
\cite{ksw,tulin_etal2021,yangyu2022}, 
have taken account of the indistinguishability of the $\chi$ particles in the 
reaction (\ref{chichi_reaction}). However, we are not
aware of published fits to observational data sets used in \cite{kaplinghat_tulin_yu_prl2016} and 
\cite{valli_yu} using the SIDM velocity-dependent cross sections in the Born
regime that compare fitted values obtained from calculations using 
cross sections with $t$-channel and $u$-channel terms included, as
compared to cross sections that only included
the $t$-channel terms. We thus proceed with these comparative fits.  

% =================================================================
%
\begin{figure*}[ht!]
	\begin{center}
		%fig. 3a
		\begin{subfigure}{0.48\textwidth}
		        \centering
			\includegraphics[width=\textwidth]{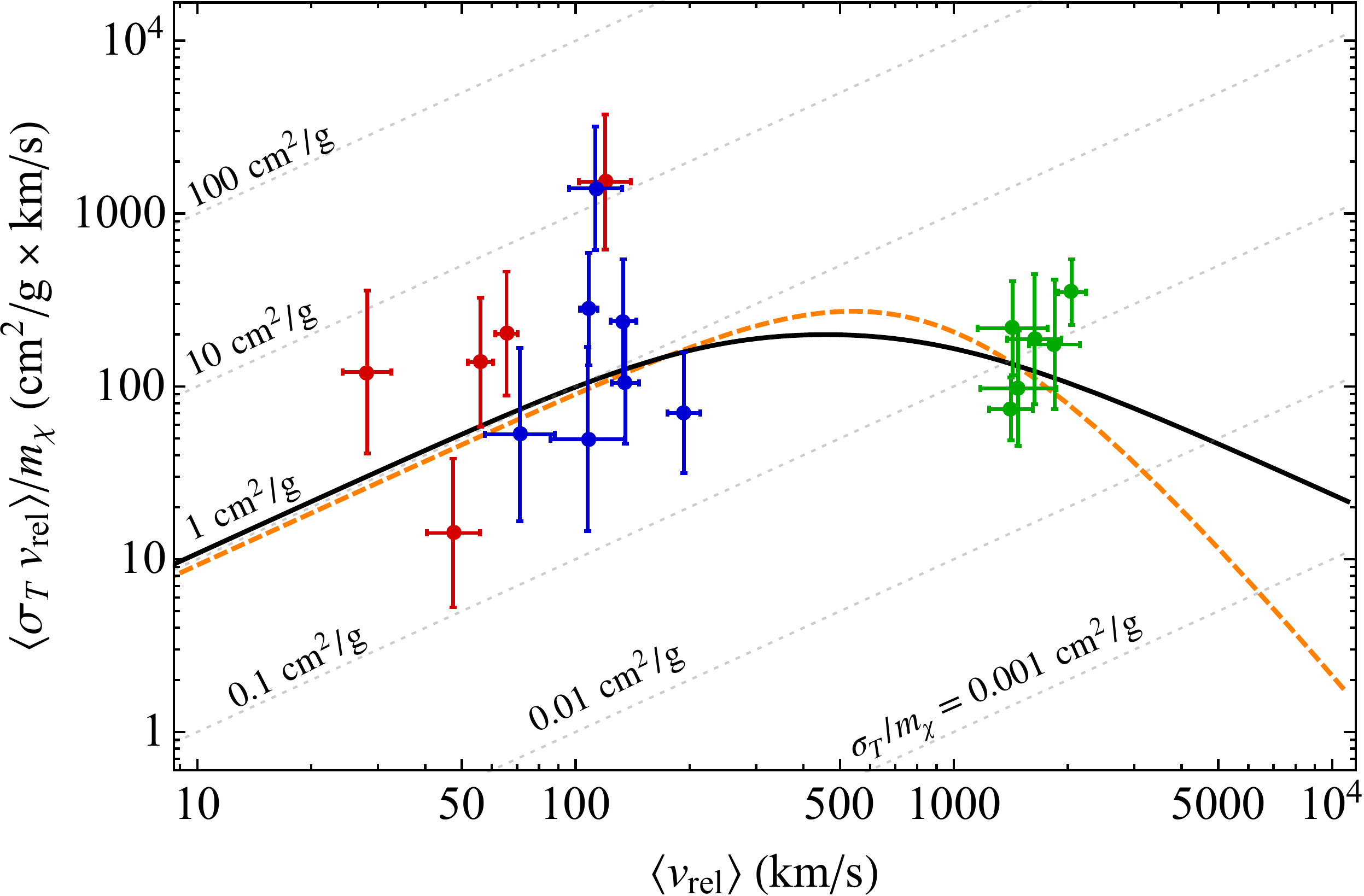}
			\subcaption{}
			\label{sigma_transfer_plot}
		\end{subfigure}
		\hspace{0.3cm}
		%fig. 3b
		\begin{subfigure}{0.48\textwidth}
			\centering
			\includegraphics[width=\textwidth]{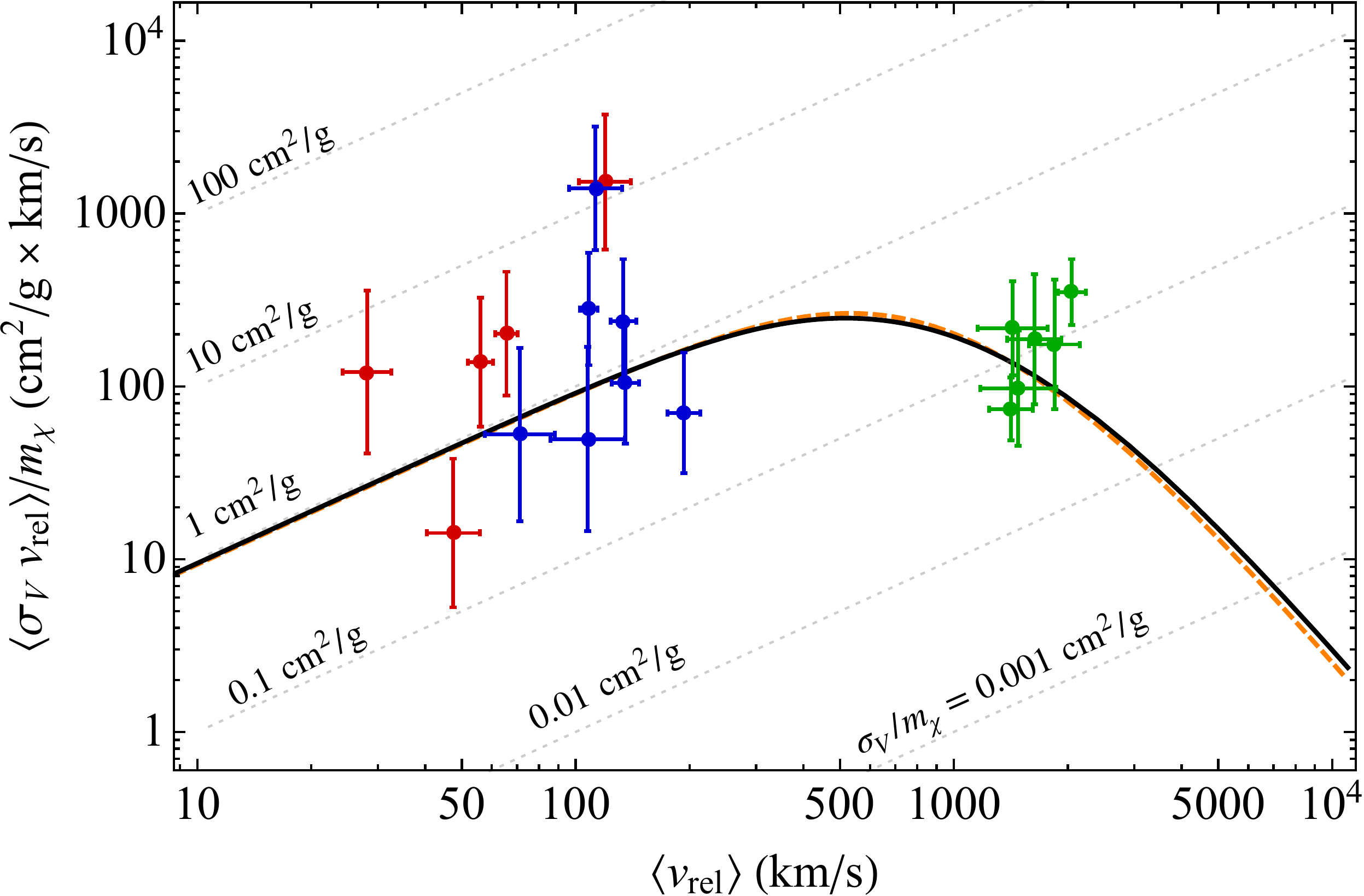}
			\subcaption{}
			\label{sigma_viscosity_plot}
		\end{subfigure}
	\end{center}
	\caption{\footnotesize Fits (black curves) of our (a)
          $\sigma_T/m_\chi$ and (b) $\sigma_V/m_\chi$ to observational
          data, where $\sigma_T$ and $\sigma_V$ are given in
          Eqs. (\ref{sigma}) and Eq. (\ref{sigma_v}).  The data are
          from dwarfs (red), LSB galaxies (blue), and galaxy clusters
          (green) as in Ref. \cite{kaplinghat_tulin_yu_prl2016}.  For
          comparison, fits to this data set with $\sigma_T$ and
          $\sigma_V$ from Eqs. (\ref{sigma_t_lit}) and
          (\ref{sigma_v_lit}), based on Eq. (\ref{dsig_lit}), are
          shown as the dashed orange curves.  Note that
          Ref. \cite{yangyu2022} finds that $\sigma_V$ provides a
          better description of thermalization effects due to SIDM
          scattering than $\sigma_T$.}
\label{fit_plots}
\end{figure*}
%

%=================================================================

\begin{figure*}[ht!]
	\begin{center}
		\begin{subfigure}{0.48\textwidth}
			\centering
		\includegraphics[width=\textwidth]{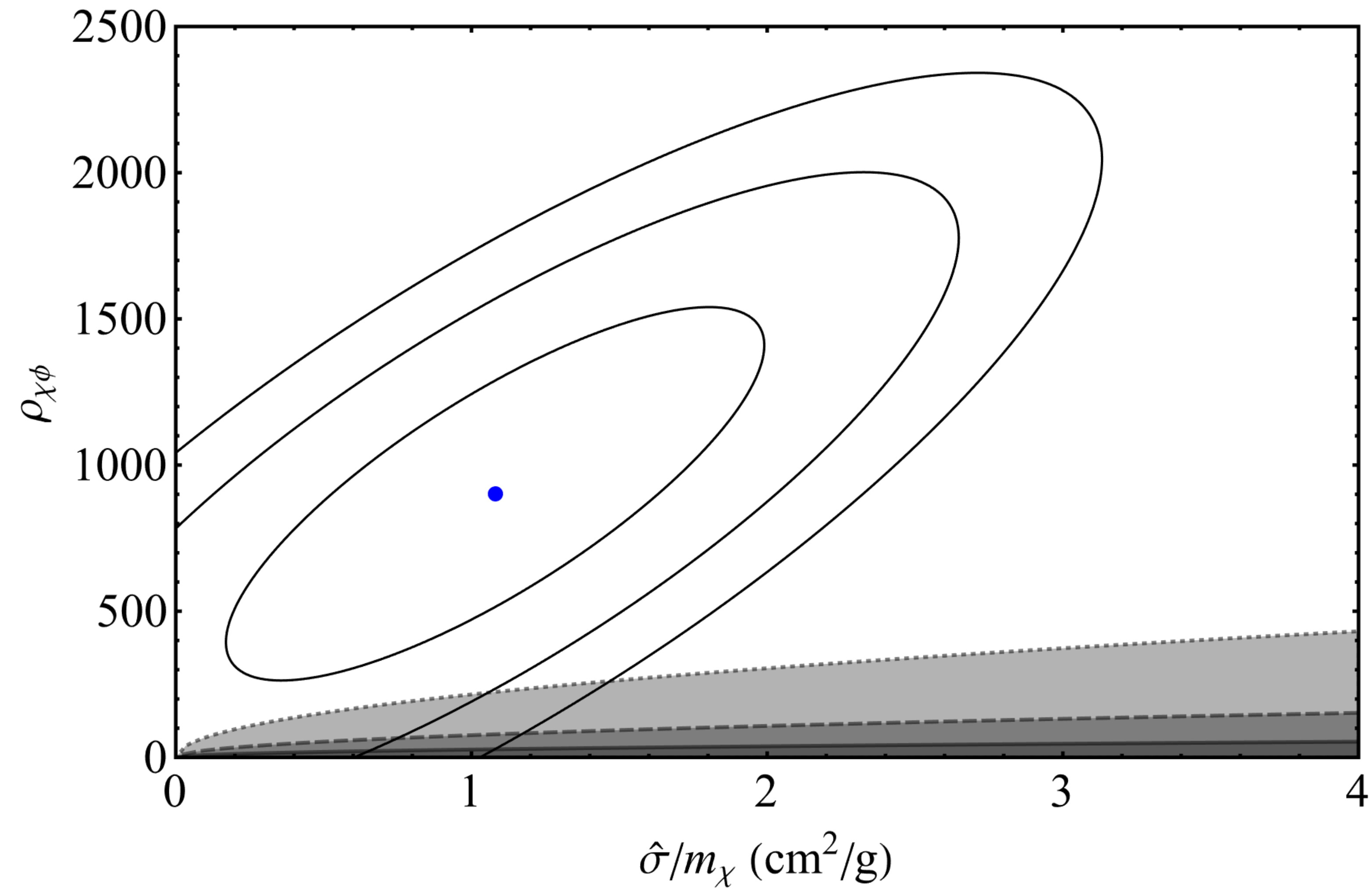}
			\subcaption{}
			\label{tplusu_transfer_para_fig}
		\end{subfigure}
		\hspace{0.3cm}
		%fig. 3d
		\begin{subfigure}{0.48\textwidth}
			\centering
			\includegraphics[width=\textwidth]{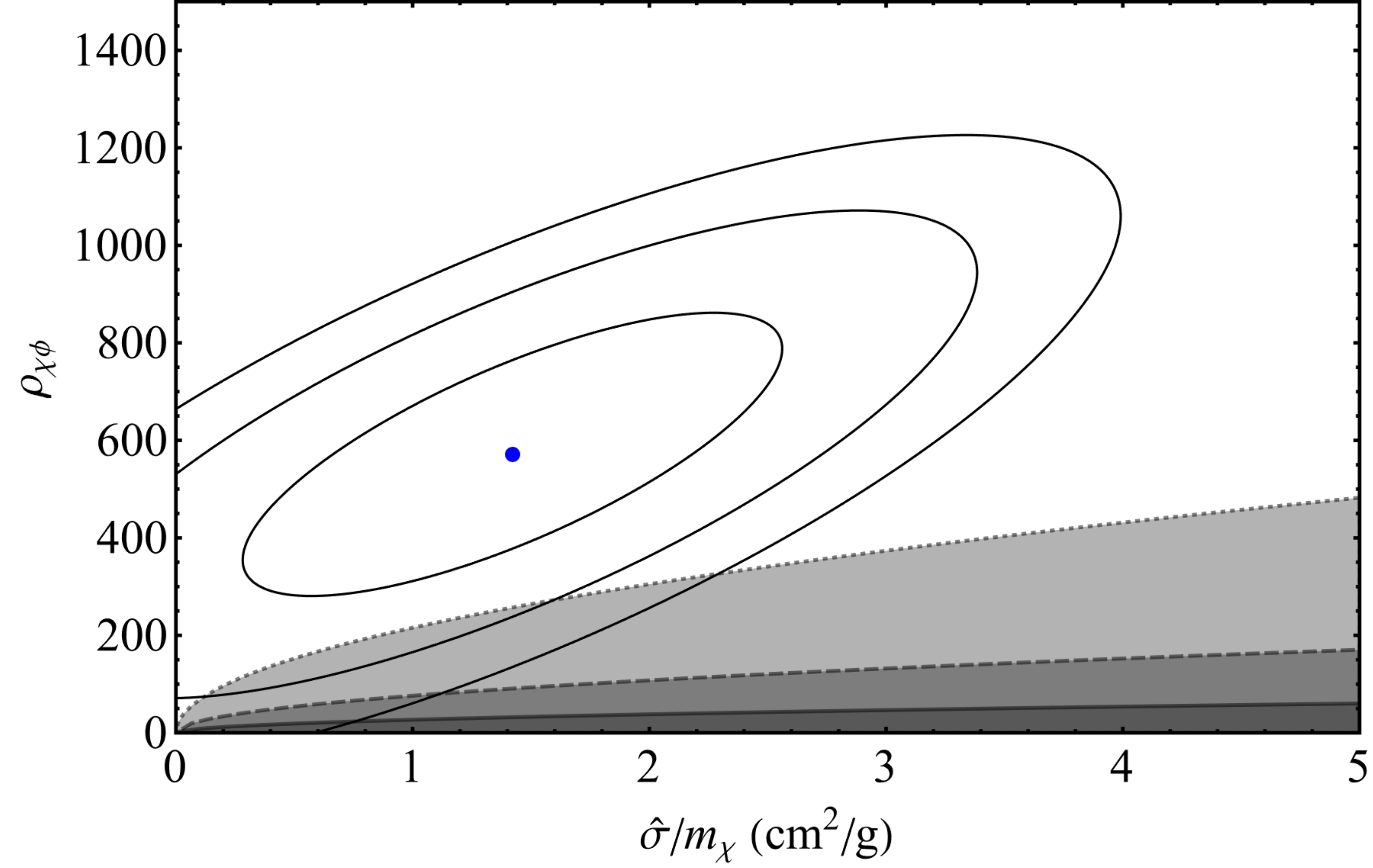}
			\subcaption{}
			\label{tplusu_vis_para_fig}
		\end{subfigure}
	\end{center}
	\caption{\footnotesize Plots showing 68\%, 95\%, and 99\%
          confidence-level (CL) contours (the inner, middle, and outer curves,
          respectively) for the fitting parameters $\rho_{\chi\xi}$ and (a)
          $\hat\sigma_T/m_\chi$, (b) $\hat\sigma_V/m_\chi$.  
          In each figure, the
          blue point denotes the respective best-fit point. The curves show the
          boundaries between the parameter regions where our Born calculation
          is applicable (light regions) and inapplicable (dark regions) 
          for the illustrative mass values $m_\chi=1, \ 2, \ 4$ GeV,
          with the larger
          excluded regions applying for larger $m_\chi$.  See text and
          caption to Fig. \ref{fit_plots} for further details.}
	\label{parameter_cl_plots}
\end{figure*}
%

%===============================================================

% ================================================================

\begin{figure*}[ht!]
	\begin{center}
		%fig. 3a
		\begin{subfigure}{0.48\textwidth}
			\centering
			\includegraphics[width=\textwidth]{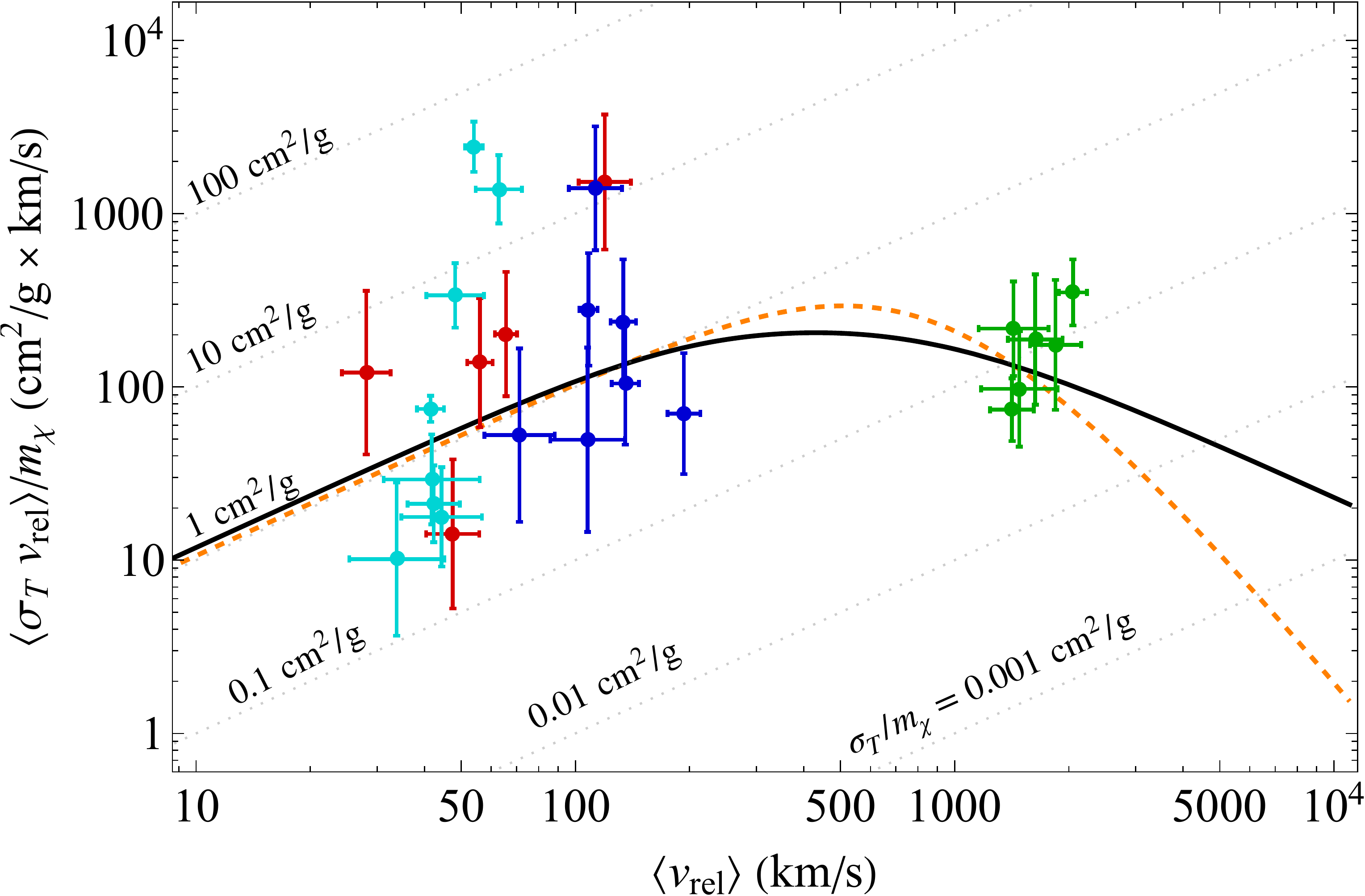}
			\subcaption{}
			\label{sigma_transfer_plot2}
		\end{subfigure}
		\hspace{0.3cm}
		%fig. 3b
		\begin{subfigure}{0.48\textwidth}
			\centering
			\includegraphics[width=\textwidth]{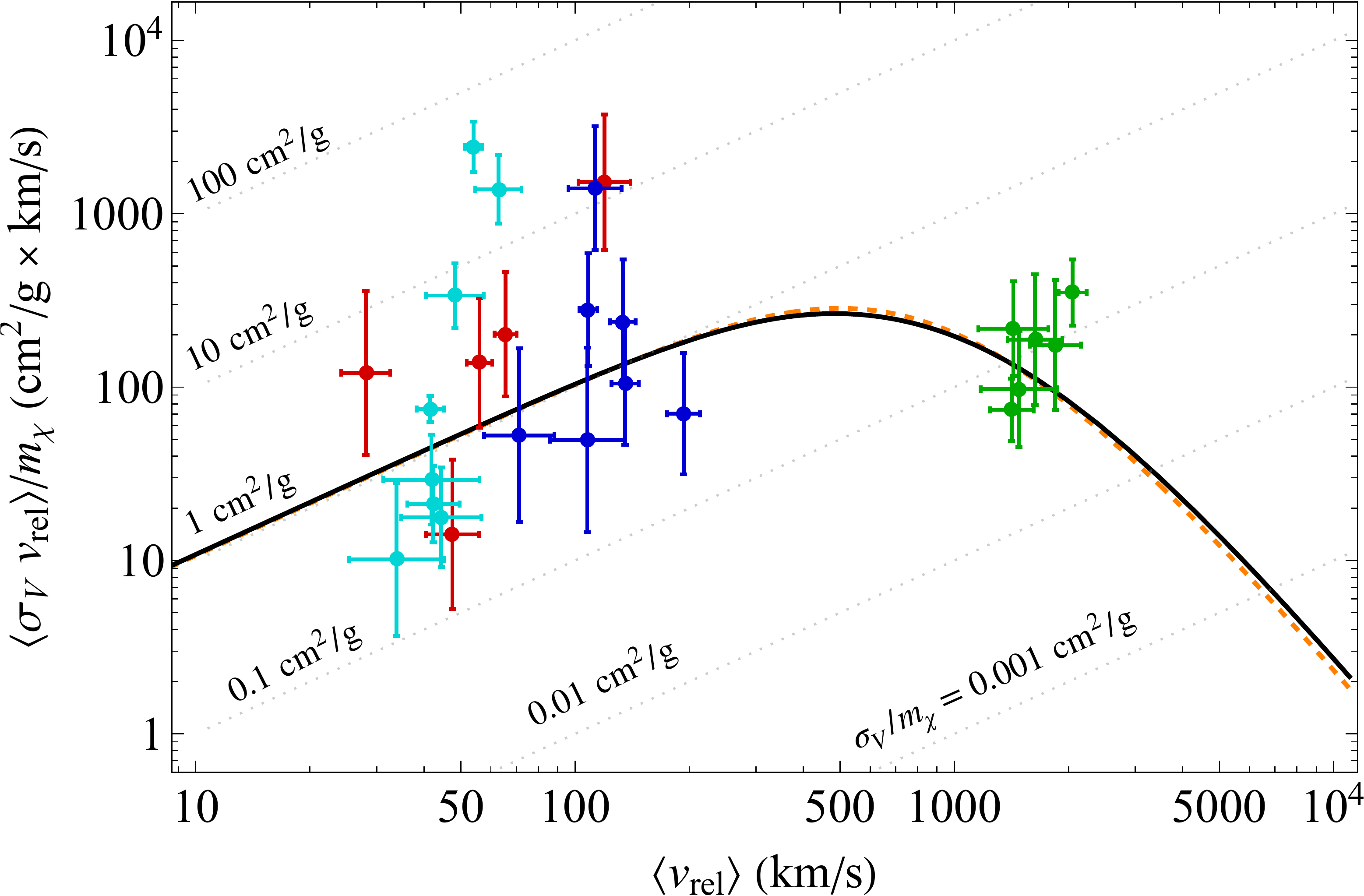}
			\subcaption{}
			\label{sigma_viscosity_plot2}
		\end{subfigure}
	\end{center}
	\caption{\footnotesize Fit (black curve) of our (a)
          $\sigma_T/m_\chi$ and (b) $\sigma_V/m_\chi$ to observational
          data augmented by inclusion of Milky Way dwarfs, where
          $\sigma_T$ and $\sigma_V$ are given in Eqs. (\ref{sigma})
          and Eq. (\ref{sigma_v}).  The data are from field dwarfs
          (red), LSB galaxies (blue), galaxy clusters (green), and
          classical Milky Way dwarfs (cyan), the latter from
          Ref.~\cite{valli_yu}. For comparison, fits to the
          $\sigma_T/m_\chi$ and $\sigma_V/m_\chi$ to this data set,
          where $\sigma_T$ and $\sigma_V$ are given in
          Eq. (\ref{sigma_t_lit}) and Eq. (\ref{sigma_v_lit}) based on
          Eq. (\ref{dsig_lit}), are shown as the dashed orange
          curves. Note that Ref. \cite{yangyu2022} finds that
          $\sigma_V$ provides a better description of thermalization
          effects due to SIDM scattering than $\sigma_T$.}
	\label{fit_plots2}
\end{figure*}
%

% ===================================================================

\begin{figure*}[ht!]
	\begin{center}
		\begin{subfigure}{0.48\textwidth}
			\centering
			\includegraphics[width=\textwidth]{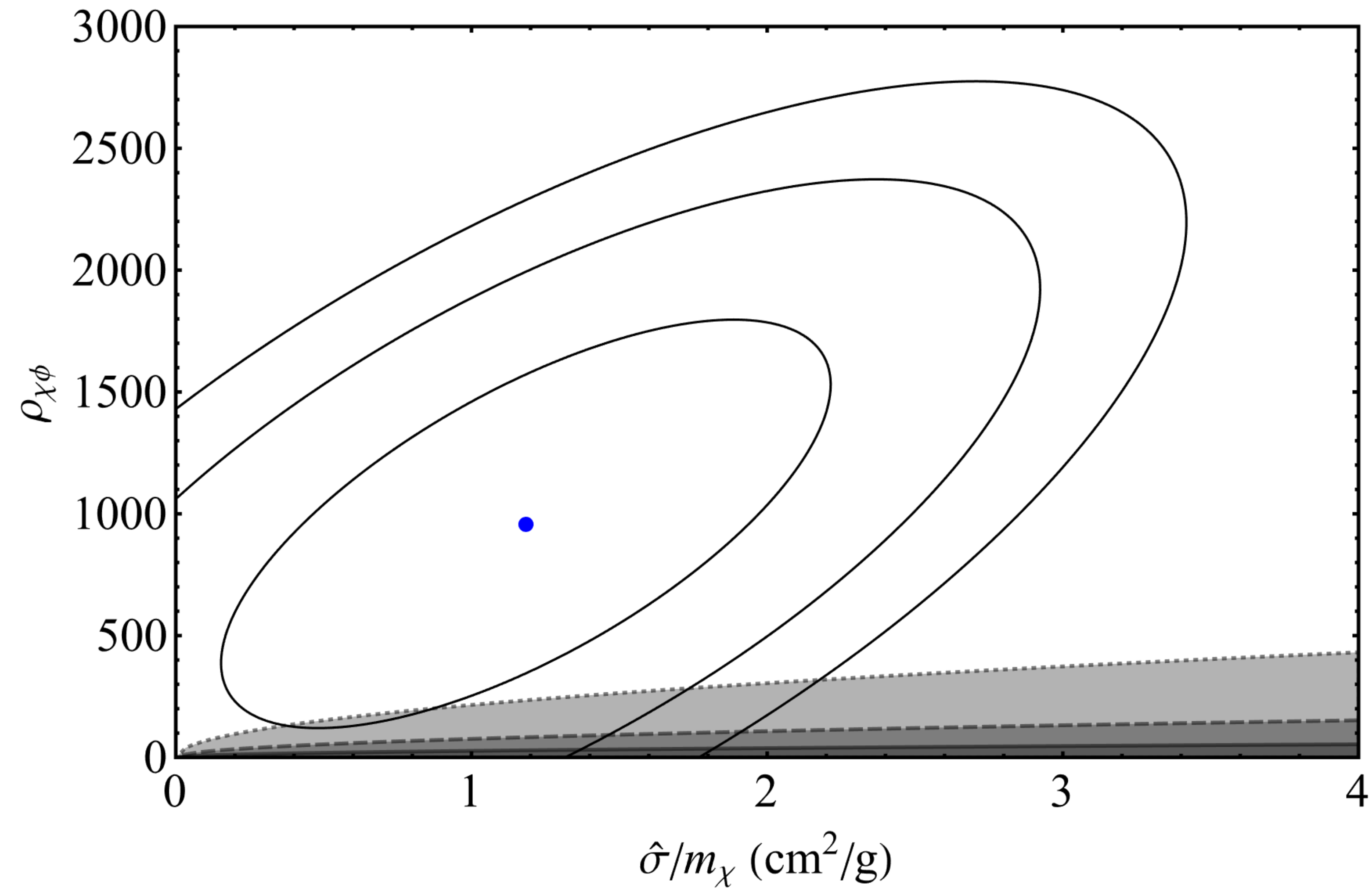}
			\subcaption{}
			\label{tplusu_transfer_para_fig2}
		\end{subfigure}
		\hspace{0.3cm}
		%fig. 3d
		\begin{subfigure}{0.48\textwidth}
			\centering
			\includegraphics[width=\textwidth]{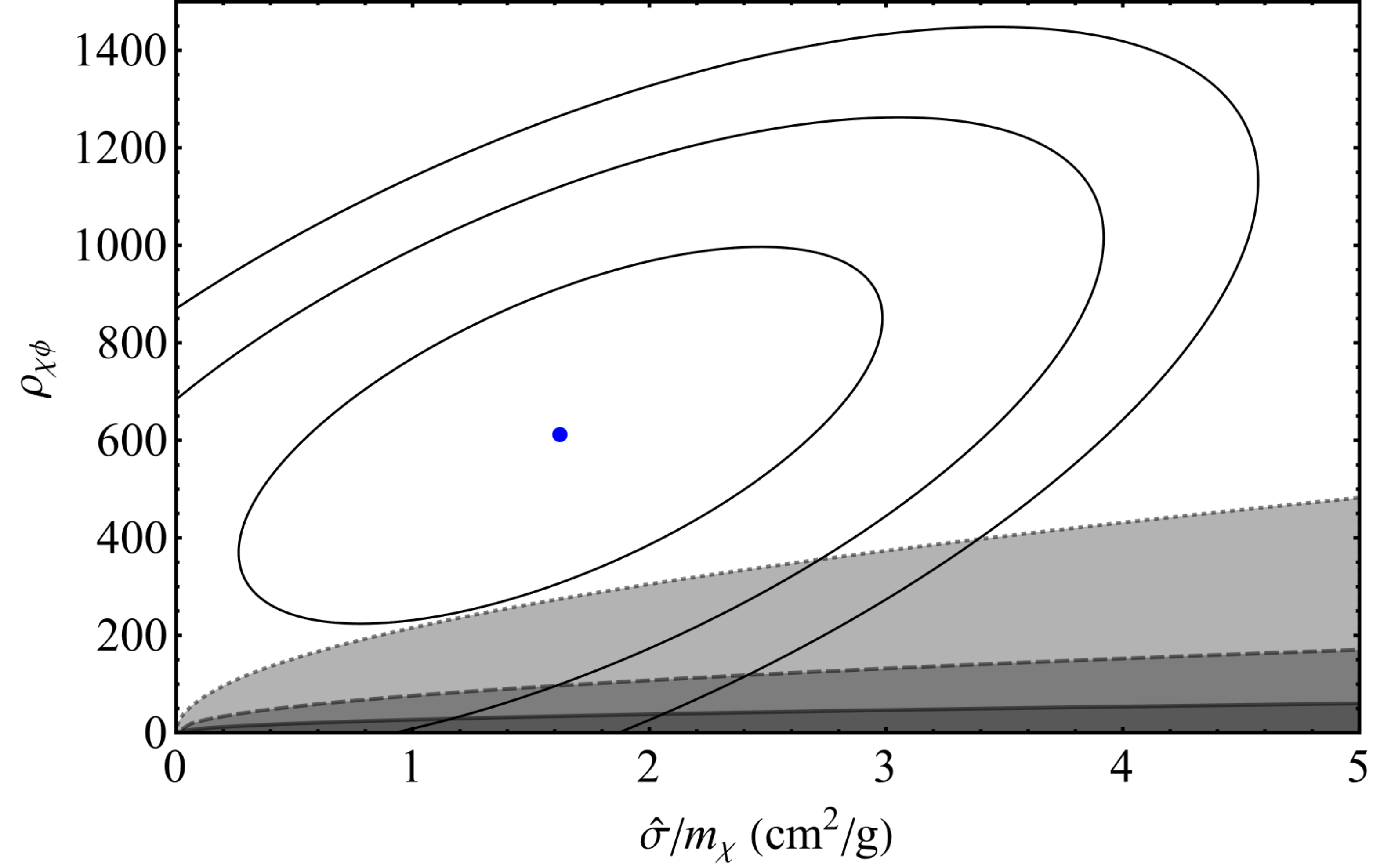}
			\subcaption{}
			\label{tplusu_vis_para_fig2}
		\end{subfigure}
	\end{center}
	\caption{\footnotesize Plots showing 68\%, 95\%, and 99\%
          confidence-level (CL) contours for the fitting parameters
          $\rho_{\chi\xi}$ and (a) $\hat\sigma_T/m_\chi$, (b)
          $\hat\sigma_V/m_\chi$ for the observational data including Milky Way
          dwarf galaxies from \cite{valli_yu}.  In each figure, the blue point
          denotes the respective best-fit point.  The curves show the
          boundaries between the parameter regions where our Born calculation
          is applicable (light regions) and inapplicable (dark regions) 
          for the illustrative mass values $m_\chi=1, \ 2, \ 4$ GeV,
          with the larger excluded regions applying for larger $m_\chi$.
          See text and caption to Fig. \ref{fit_plots2} for further details.}
	\label{parameter_cl_plots2}
\end{figure*}
%

% =====================================================================

% section 3
\section{Fits to Observational Data}
\label{fits_section}

We now address the question of how the fit to data changes when one uses the
cross sections from \cite{sidm} with both $t$-channel a
nd $u$-channel
contributions to the reaction (\ref{chichi_reaction}). In our main comparison,
for definiteness, we use the same set of data from dwarfs, low surface
brightness (LSB) galaxies, and galaxy clusters as in
\cite{kaplinghat_tulin_yu_prl2016}.  This data set includes (i) five dwarfs
from THINGS (The H I Nearby Galaxy Survey) \cite{things}, namely IC 2574, NGC
2366, Holmberg II, M81 dwB, and DDO 154, indicated in red in
Figs. \ref{fit_plots}(a,b); (ii) seven LSB
galaxies from \cite{naray_mcgaugh_deblok}, namely UGC 4325, F563-V2, F563-1,
F568-3, UGC 5750, F583-4, and F583-1, indicated in blue in
Figs. \ref{fit_plots}(a,b); and (iii) six
relaxed galaxy clusters from \cite{newman1,newman2}, namely MS2137, A611, A963,
A2537, A2667, and A2390, indicated in green in Figs.~\ref{fit_plots}(a,b). 
Of these, the galaxy clusters are at distances from us 
of several hundred Mpc, the LSB galaxies are at distances $\sim O(10)$ Mpc, and
the dwarfs are at distances of approximately 3 to 5 Mpc.  This choice of dwarf
galaxies considered in \cite{kaplinghat_tulin_yu_prl2016} has the merit that
these are ``field'' dwarfs located sufficiently far from the Milky
Way that they are less subject to the complication of environmental effects,
such as possible tidal stripping, than dwarfs closer to the Milky Way, such as
the so-called classical dwarfs \cite{valli_yu}.

We carry out fits to $\sigma_T/m_\chi$ and $\sigma_V/m_\chi$. For a given
$v_{\rm rel}$, these cross sections depend on two parameters, which are thus
determined by the fits to the data. Since the small-$\beta_{\rm rel}$ limit
(which is also the $r \to 0$ limit for a given ratio $m_\chi/m_\xi$) of our
$\sigma_T$ in Eq. (\ref{sigma}) is
\beq
\lim_{r \to 0} \sigma_T = 2\pi \sigma_0 \ , 
\label{sigma_T_r0}
\eeq
so that 
\beq
\lim_{r \to 0} \frac{\sigma_T}{m_\chi} = 
\frac{2\pi \alpha_\chi^2 m_\chi}{m_\xi^4} \ , 
\label{sigma_T_over_m_r0}\eeq
we take one fitting parameter to be
\beq
\frac{\hat \sigma}{m_\chi} \equiv \frac{2\pi \alpha_\chi^2 m_\chi}{m_\xi^4} \ .
\label{sigma_hat_over_mchi}
\eeq
Here we use the symbol $\hat \sigma$ to indicate that this is a fitting
parameter.  Note also that in the same limit, our $\sigma_V$ in
Eq. (\ref{sigma_v}) has the value 
\beq
\lim_{r \to 0} \sigma_V = \frac{2}{3} \lim_{r \to 0} \sigma_T = 
\frac{4\pi}{3} \sigma_0 \ . 
\label{sigma_V_r0}
\eeq
The other fitting parameter is the ratio 
\beq
\quad  \rho_{\chi\xi}  \equiv \frac{m_\chi}{m_\xi} \ , 
\label{fit_parameters}
\eeq
which, for a given $v_{\rm rel}$, determines $r$.  For this comparison we we
utilize the same central values and estimated error bars as in Fig. 1 of
\cite{kaplinghat_tulin_yu_prl2016}.  We use a $\chi^2$ fitting procedure in the
{\tt NonlinearModelFit} routine in Mathematica, with our formulas (\ref{sigma})
and (\ref{sigma_v}) for the transfer and viscosity cross-sections.

Our results are shown as the black curves in Figs.  
\ref{fit_plots2}(a,b). From the $\sigma_T$ fit, we find 
\beq
\frac{\hat\sigma}{m_\chi} = 1.1 \pm 0.6 \ {\rm cm}^2/{\rm g} \ , \quad 
\rho_{\chi\phi}  = (0.90 \pm 0.41) \times 10^3 \ . 
\label{sigma_transfer_fit_parameters}
\eeq
From the $\sigma_V$ fit, we find the values 
\beq
\frac{\hat\sigma}{m_\chi} = 1.4 \pm 0.7 \ {\rm cm}^2/{\rm g} \ , \quad 
\rho_{\chi\phi}  = (0.57 \pm 0.19) \times 10^3 \ . 
\label{sigma_viscosity_fit_parameters}
\eeq
The confidence-level contours for these fits are shown in Figs.  
\ref{parameter_cl_plots}(a,b). The values of $\chi^2/{\rm DOF}$ are
$1.40$, and $1.35$ for the transfer and viscosity cross-section fit
respectively. Here, the number of degrees of freedom (DOF) is equal to 
the number of data points minus the number of free parameters
$= 18-2 = 16$.  These $\chi^2/{\rm DOF}$ values
indicate that the fits are reasonably good. As a check on this fitting
procedure, we also performed a orthogonal distance regression (ODR) in Python,
utilizing the {\tt scipy.odr} module, where instead of the least square
distances, the ODR minimizes the sum of squared perpendicular distances of each
data point from the fitted curve. Within the uncertainties in the fitted
parameters, we find good agreement between these two fitting methods.
We have checked that these fitted values are consistent with our perturbative
calculation, i.e., that they are in the Born regime.  As discussed in Appendix
A of \cite{sidm}, the condition for the validity of the Born approximation is
that the the quantity 
$\alpha_\chi m_\chi/m_\xi \equiv \alpha_\chi \rho_{\chi\xi}$ should be small
compared with unity. This quantity can be expressed in terms of our
fitting parameters $\hat\sigma/m_\chi$ and $\rho_{\chi\xi}$, together with 
$m_\chi$, as
\beq
\alpha_\chi \rho_{\chi\xi} =
\bigg [ \bigg ( \frac{\hat\sigma/m_\chi}{2\pi \rho_{\chi\xi}^2} \bigg ) \,
m_\chi^3 \bigg ]^{1/2} \ . 
\eeq
We have plotted curves along which $\alpha_\chi \rho_{\chi\xi}=1$ for three
illustrative values, $m_\chi=1, \ 2, \ 4$ GeV in Figs.
\ref{parameter_cl_plots} and \ref{parameter_cl_plots2}. The dark regions are
outside the Born regime.  Thus, for values of $m_\chi$ of a few GeV, as is
plausible in asymmetric DM models, our fitted values are consistent with the
Born approximation that we use.  A similar comment applies to our other fits
given in this paper. For $m_\chi$ values outside this range, our Born analysis
would not apply.  We emphasize that our analysis only applies for model
parameters that are in the Born regime and that also lead to sufficient
depletion of the symmetric dark matter density. Our results do not exclude
$m_\chi$ values larger than this range of a few GeV, and the values of
$\hat\sigma/m_\chi$ and $\rho_{\chi\xi}$ could be different in these cases. For
example, Ref. \cite{tulin_etal2021} obtained fits to observational data in a
light scalar mediator model with $\alpha_\chi=0.5$, $m_\chi=190$ GeV, and
$m_\phi=3$ MeV. Since the quantity $\alpha_\chi \rho_{\chi\xi} = 3.2 \times
10^4$ in this case (far outside the Born regime), this shows that it is
possible to fit data in a light-mediator SIDM model with quite different values
of model parameters.

To address the question posed in this paper, we have carried out corresponding
fits to these data using the cross sections $\sigma_{T,{\rm lit.}}$ and
$\sigma_{V,{\rm lit.}}$ obtained by including only the $t$-channel
contributions. These are shown as the dashed orange curves in
Figs. \ref{fit_plots}(a) and \ref{fit_plots}(b), respectively. The
small-$\beta_{\rm rel}$ limits (i.e., $r \to 0$ limits, for a fixed ratio
$m_\chi/m_\xi$) of $\sigma_{T,{\rm lit.}}$ and $\sigma_{V,{\rm
    lit.}}$ are given by 
\beq
\lim_{r \to 0} \sigma_{T,{\rm lit.}} = 4\pi \sigma_0 
\label{sigma_T_lit_r0}
\eeq
and
\beq
\lim_{r \to 0} \sigma_{V,{\rm lit.}} = \frac{8\pi}{3}\sigma_0 \ , 
\label{sigma_V_lit_r0}
\eeq
which are twice as large as the corresponding small-$\beta_{\rm rel}$ limits of
our cross sections.  We thus expect that the fitted value of
$\hat\sigma/m_\chi$ using the results (\ref{sigma_t_lit}) and
(\ref{sigma_v_lit}) will be roughly half the value obtained with the correct
formula, and this is borne out by our analysis.  With the $\sigma_{T,{\rm
    lit.}}$ in Eq. (\ref{sigma_t_lit}), our fit to these data, shown in
Fig. \ref{fit_plots}(a), yields
\beq
\frac{\hat\sigma}{m_\chi} = 0.45 \pm 0.25 \ {\rm cm}^2/{\rm g} \ , \quad 
\rho_{\chi\phi}  = (0.42 \pm 0.12) \times 10^3 . 
\label{sigma_t_lit_fit_parameters}
\eeq
With the $\sigma_{V,{\rm lit.}}$ in Eq. (\ref{sigma_v_lit}), our fit 
to the data, as shown by the dashed orange
curve in Fig. \ref{fit_plots}(b), gives
\beq
\frac{\hat\sigma}{m_\chi} = 0.70 \pm 0.35 \ {\rm cm}^2/{\rm g} \ , \quad 
\rho_{\chi\phi}  = (0.51 \pm 0.16) \times 10^3 \ . 
\label{sigma_v_lit_fit_parameters}
\eeq
For both of these fits, the reduced $\chi^2/{\rm DOF} = 1.35$.  Again, these
$\chi^2/{\rm DOF}$ values show that these are reasonably good fits, albeit with
the above-mentioned differences in the values of the fitting parameters.  
Our curves are also consistent with the results of \cite{tulin_velocity}, which
used modelling methods and simulation taking into account both baryon effects
and SIDM. 

As we pointed out in \cite{sidm}, for values of $v_{\rm rel} \gsim 2
\times 10^3$ km/s typical of galaxy clusters, our $\sigma_T$ is
considerably larger than the result (\ref{sigma_t_lit}), and this is
evident in the deviation between the black curve and orange dashed
curve in Fig. \ref{fit_plots}(a). However, with the current data set,
this deviation does not have much effect on our fit, since the galaxy
clusters have $v_{\rm rel}$ values between $\sim 1 \times 10^3$ km/s
and $2 \times 10^3$ km/s. It is also possible that $\sigma_T$
overestimates the effects of SIDM self-scattering when one includes
both $t$-channel and $u$-channel contributions.

In \cite{sidm} we calculated that
$\sigma_V/\sigma_V^{(t)} = 1 + (1/10)r^2 + O(r^3)$ for $r \ll 1$ and
$\sigma_V/\sigma_V^{(t)} = 2 = (1/\ln r)$ for $r \gg 1$ (see
Eqs. (4.44) and (4.45) of \cite{sidm}). Thus, not only for small $r$,
but also for $r \gsim 1$, $\sigma_V^{(t)}$ and $\sigma_V$ have rather
similar functional dependence on $r$ and hence also on $v_{\rm rel}$.
This similarity property, shown in \cite{sidm}, is again evident in
Fig.  \ref{fit_plots}(b). However, since $\sigma_{V,{\rm lit.}}$ is
twice as large as our $\sigma_V^{(t)}$ from (Eq. (4.36) of
\cite{sidm}), a fit of $\sigma_V,{\rm lit.}$ is expected to yield a
value of the fitting parameter $\hat \sigma/m_\chi$ that is
approximately half as large as a value obtained from a fit using our
$\sigma_V$, and this expectation is again borne out by our results.

It is also of interest to compare the values of $\sigma_T/m_\chi$ and
$\sigma_V/m_\chi$ from our fit to data with the illustrative set of
values that we used in \cite{sidm}.  In that paper we utilized the
input values $m_\chi = 5$ GeV, $m_\xi = 5$ MeV, and $\alpha_\chi = 3
\times 10^{-4}$. This set of parameters yielded the values 0.99, 0.89,
0.13 in units of cm$^2$/g for $\sigma_T/m_\chi$ and the values 0.66,
0.59, and 0.030 in units of cm$^2$/g for $\sigma_V/m_\chi$, for
$v_{\rm rel} = 10, \ 10^2, \ 10^3$ km/sec, respectively (see Table 1
of \cite{sidm}). Evidently, these value from \cite{sidm} are close to
results of the actual fit to observational data that we have carried
out here.  The value of $\rho_{\chi\xi}$ from our fit to
$\sigma_T/m_\chi$ is in very good agreement, to within the
uncertainty, with the value $\rho_{\chi\xi}=10^3$ for the illustrative
set in \cite{sidm}, while the value of $\rho_{\chi\xi}$ from our fit
to $\sigma_V/m_\chi$ is slightly smaller than the above-mentioned
illustrative value in \cite{sidm}, but is well within the Born regime
shown in Fig. 3 of \cite{sidm}.

It is also worthwhile to consider fits to a larger set of astronomical
data.  For this purpose, we have carried out the analogous fits of
$\sigma_T/m_\chi$ and $\sigma_V/m_\chi$ to a data set consisting of
the above objects (field dwarfs, LSB galaxies, and galaxy clusters)
considered in \cite{kaplinghat_tulin_yu_prl2016} augmented by the
classical Milky Way dwarf spheroidal (dSph) galaxies, as analyzed in
\cite{valli_yu}, namely Ursa Minor, Draco, Sculptor, Sextans, Carina,
Fornax, Leo I, and Leo II), with distances ranging from 76 kpc (for
Draco) to 254 kpc (for Leo I) \cite{des2020}. Since these classical
dSph galaxies are closer to the disk of the Milky Way than the field
dwarfs that were fitted in \cite{kaplinghat_tulin_yu_prl2016}, they
are more susceptible to environmental effects due to the Milky Way,
including tidal stripping, than these more distant field dwarfs, as was
cautioned in \cite{valli_yu} and has been studied further in
\cite{slone_evolution,bullock_kaplinghat_valli2022}. 

With these caveats in mind, we show our results in Figs.
\ref{fit_plots2}(a,b). The corresponding confidence-level contour plots are
presented in Figs. \ref{parameter_cl_plots2}(a,b). As is evident, there is
considerable scatter in the data in this larger data set.  The best fit
parameters that we find for this data set are
\beq 
\frac{\hat\sigma}{m_\chi} = 1.2 \pm 0.7 \ {\rm cm}^2/{\rm g} \ , \quad
\rho_{\chi\phi} = (0.96 \pm 0.54) \times 10^3
\label{sigma_t_vy_fit_parameters}
\eeq
and
\beq
\frac{\hat\sigma}{m_\chi} = 1.6 \pm 0.9 \ {\rm cm}^2/{\rm g} \ , \quad
\rho_{\chi\phi}  = (0.61 \pm 0.25) \times 10^3 \ . 
\label{sigma_v_vy_fit_parameters}
\eeq
The corresponding values of $\chi^2/{\rm DOF}$ for the fit to this
larger data set are $2.92$ and, $2.76$ for the transfer and viscosity
cross sections, respectively. The increase in $\chi^2/{\rm DOF}$,
i.e., reduction in the goodness of fit, is presumably associated with
the greater scatter (diversity) in the Milky Way dSph data set from
\cite{valli_yu}.  This scatter may be understood better as a result of
improved modelling of these Milky Way dwarfs
\cite{slone_evolution,bullock_kaplinghat_valli2022} (and references
therein).

With the $\sigma_{T,{\rm lit.}}$ and 
$\sigma_{V,{\rm lit}}$ in Eqs. (\ref{sigma_t_lit}) and
(\ref{sigma_v_lit}) used for the theoretical model, the fit parameters 
are 
\beq
\frac{\hat\sigma}{m_\chi} = 0.5 \pm 0.3 \ {\rm cm}^2/{\rm g} \ , \quad
\rho_{\chi\phi}  = (0.44 \pm 0.17) \times 10^3
\label{sigma_t_vy_lit_fit_parameters}
\eeq
and
\beq
\frac{\hat\sigma}{m_\chi} = 0.8 \pm 0.4 \ {\rm cm}^2/{\rm g} \ , \quad
\rho_{\chi\phi}  = (0.54 \pm 0.21) \times 10^3 \ . 
\label{sigma_v_vy_lit_fit_parameters}
\eeq
The reduced $\chi^2/{\rm DOF}$ values for these fits are $2.73$ and $2.74$,
respectively. Note that for the larger data set, ${\rm DOF} = 26 - 2 = 24$.
In future work, one could further enlarge the data sets for SIDM fits. 
Indeed, several dedicated observational surveys have considerably expanded 
the number of Milky Way dwarf satellites in recent years, 
in particular, with the detection of ultra-faint dwarfs 
\cite{des2020,simon_review}.

It is useful to consider a rescaling of parameters for the cross
sections based on inclusion of only the $t$-channel contribution for
$\sigma_{T,{\rm lit.}}$, Eq. (\ref{sigma_t_lit}), and for
$\sigma_{V,{\rm lit.}}$, Eq.  (\ref{sigma_v_lit}), that minimizes the
respective deviations from the cross sections calculated with
inclusion of both $t$-channel and $u$-channel contributions (and their
interference), given by Eqs. (\ref{sigma}) for $\sigma_T$ and by
Eq. (\ref{sigma_v}) for $\sigma_V$. For this purpose, we use our fit
to the observational data in \cite{kaplinghat_tulin_yu_prl2016}. We
find that an overall rescaling of $\sigma_0 \to (1/2)\sigma_0$, and $r
\to 0.75 r$ for the fit with $\sigma_{V,{\rm lit.}}$ in
Eq. (\ref{sigma_v_lit}) numerically minimizes its deviation from the
results obtained with $\sigma_V$ in Eq. (\ref{sigma_v}). Here the
factor of 1/2 accounts for the identical final-state particles. Note
that this rescaling is consistent with the best fit obtained in
Eqs. (3.7), and (3.12). Furthermore, from Eq. (2.1), if one fixes
$m_\chi$, then the above rescaling implies a slight shift in the
underlying physical parameters, namely $m_\xi \to 1.15 m_\xi$, and
$\alpha_\chi \to 0.94 \alpha_\chi$. In order to minimize the deviation
of cross-sections in Eq.~(\ref{sigma_t_lit}), and (\ref{sigma_v}), we
can rescale $\sigma_0 \to \sigma_0/3$, and $r \to 0.45r$. The
rescaling $\sigma_0 \to \sigma_0/3$ can be understood as the
combination of the factor $2/3$ coming from the different weights in
the definition of transfer and viscosity cross-section and the factor
$1/2$ to account for the identical final state particles. Similarly,
this implies a slight rescaling of $\alpha_\chi \to 1.28 \alpha_\chi$,
and $m_\xi \to 1.49 m_\xi$. This agrees with our best fits in
Eqs.~(\ref{sigma_viscosity_fit_parameters}) and
(\ref{sigma_t_lit_fit_parameters}).  These rescaling can be useful to
estimate the conversion of results in the literature, based on
inclusion of only $t$-channel contributions, to results from
calculations of cross sections based on inclusion of both the
$t$-channel and $u$-channel terms in the Born regime.

% ================================================================

% section 4
\section{Conclusions}
\label{conculsion_section}

In conclusion, self-interacting dark matter models provide an
appealing way to avoid problems encountered in pure cold dark matter
simulations lacking baryon feedback.  In this paper we have continued
our study of an asymmetric dark matter model with self-interactions,
in the Born parameter regime.  In \cite{sidm} we calculated
differential and integrated cross sections that take into account both
$t$-channel and $u$-channel contributions to the scattering of
identical dark matter fermions via exchange of a light mediator
particle. Our work in \cite{sidm} was motivated in part by the 
previous use of cross section formulas that included only $t$-channel
contributions in fits to astronomical data.  Here we have investigated
a question that arose from our analysis in \cite{sidm}, namely how do
the results of these fits change when one uses cross sections that
correctly include both $t$-channel and $u$-channel contributions. Our
results for the fitting parameters $\hat \sigma/m_\chi$ and
$\rho_{\chi \xi}$ are somewhat different from the values that one
would get if one were to use only $t$-channel contributions.
Nevertheless, our broad conclusions are in agreement with previous
studies, namely that this type of self-interacting dark matter model
with a light mediator can provide a reasonably good fit to a variety
of observational data ranging from dwarfs to galaxy clusters. Further
progress in modelling and comparison of SIDM models with observational
data should shed additional light on this promising class of dark
matter models.

\begin{acknowledgments} 

  We thank Dr. Mauro Valli for valuable discussions and Prof. Hai-Bo
  Yu for useful email exchanges.  This research was supported in part
  by the U.S. National Science Foundation Grants NSF-PHY-19-15093 and
  NSF-PHY-22-15093 (R.S.) and by support from Tsung-Dao Lee Institute
  (S.G.).

\end{acknowledgments} 

% ===========================================================

% ===============================================
%\clearpage
%\nocite{*}
\bibliographystyle{apsrev4-2}
\bibliography{sidmfit}
%===============================================
\end{document}